\documentclass[prl,twocolumn]{revtex4}
\usepackage{epsfig,psfrag,amsmath,amssymb,lmodern}
\usepackage{ulem}

\def \tr {\mathop{\rm tr}\nolimits}

\def \e  {\mathop{\rm e}\nolimits}
\newcommand\lr[1]{{\left({#1}\right)}}

\newcommand \ket [1] {|{#1}\rangle}
\newcommand \bra [1] {\langle {#1}|}
\newcommand\re[1]{(\ref{#1})}

\newcommand \vev [1] {\langle{#1}\rangle}
\newcommand{\ft}[2]{{\textstyle\frac{#1}{#2}}}

\def \be  {\begin{equation}}
\def \ee  {\end{equation}}
\def \ba  {\begin{eqnarray}}
\def \ea  {\end{eqnarray}}
\def \baa {\begin{eqnarray*}}
\def \eaa {\end{eqnarray*}}
\def \bb  {\begin {thebibliography} }
\def \eb  {\end{thebibliography}}

\begin{document}
\title{
\begin{flushright} ${}$\\[-40pt] $\scriptstyle \rm CERN-PH-TH/2013-282,\ IPhT-13/264,\  LAPTH-069/13$ \\[0pt]
\end{flushright}
Energy-energy correlations in ${\cal N}=4$ SYM }
\author{A.V.~Belitsky$^{a}$,  S.~Hohenegger$^b$, G.P.~Korchemsky$^c$, E.~Sokatchev$^{b,d,e}$, A.~Zhiboedov$^f$}
\affiliation{$^a$Department of Physics, Arizona State University, Tempe, AZ 85287-1504, USA}
\affiliation{$^b$Physics Department, Theory Unit, CERN, CH -1211, Geneva 23, Switzerland}
\affiliation{$^c$Institut de Physique Th\'eorique\footnote{Unit\'e de Recherche Associ\'ee au CNRS URA 2306}, CEA Saclay, 91191 Gif-sur-Yvette Cedex, France}
\affiliation{$^d$Institut Universitaire de France, 103, bd Saint-Michel F-75005 Paris, France}
\affiliation{$^e$LAPTH\,\footnote{UMR 5108 du CNRS, associ\'ee \`a l'Universit\'e de Savoie},   Universit\'{e} de Savoie, CNRS, B.P. 110,  F-74941 Annecy-le-Vieux, France} 
\affiliation{$^f$Department of Physics, Princeton University, Princeton, NJ 08544, USA}
\begin{abstract}
We present a new  approach to computing energy-energy correlations in gauge theories that exploits their relation to correlation functions and  
bypasses the use of scattering amplitudes. We illustrate its power by calculating energy-energy correlations  in the 
maximally supersymmetric Yang-Mills theory ($\mathcal{N}=4$  SYM)  in the next-to-leading order approximation.  
\end{abstract}

\maketitle

\noindent \textbf{1. Introduction. }  
In this paper we study a particular class of observables, the so-called energy-energy correlations (EEC), which have first been 
introduced in the context of QCD for the process of electron-positron annihilation into hadrons. EEC played a crucial 
role in precision tests of QCD as the theory of strong interactions. EEC measures the differential
angular distribution of the energy that flows through two calorimeters separated by the relative angle $\chi$.
It is defined \cite{Basham:1978bw} as the energy-weighted sum over pairs of particles produced in the final states of 
$\e^+\e^-\to V\to a+b+ \text{everything}$  (with $V$ being a virtual photon $\gamma^*$ or a $Z^0$ boson)
\begin{align}\label{EEC}
{\rm EEC} = \sum_{a,b}   \int \! d\sigma_{V\to {a+b}+X} {E_a E_b\over Q^2 \sigma_{\rm tot}} \delta(\cos\theta_{ab}-\cos\chi).
\end{align}
Here $Q$ is the total energy in the center-of-mass frame, $E_a$ and $E_b$ are the energies of the detected particles, $\theta_{ab}=\chi$ 
is the relative angle between their spatial momenta, $\sigma_{\rm tot}$ is the total cross-section such that $\int_{-1}^{1} d \cos \chi\, {\rm EEC} = 1$.

Being an infrared safe quantity, EEC is insensitive to long-distance dynamics and can be 
computed in perturbative QCD as a series in powers of the coupling $\alpha_s(Q)$. Nonperturbative hadronization effects modify
EEC at the level of corrections suppressed by powers of $1/Q$.   For $0<\chi<\pi$,   EEC   receives
contributions from final states with three or more particles and its perturbative expansion starts at order $O(\alpha_s)$.
The leading order effect was computed using conventional techniques \cite{Basham:1978bw,LEP2}
\begin{align} \label{1loopQCD}
{\rm EEC}_{\rm \scriptscriptstyle QCD}  = \frac{\alpha_s C_F}{4\pi z^3} 
\bigg[
& \lr{{z\over 1-z} - \frac{9 -15 z + 3z^2}{z^2}}
\ln{1\over 1-z}
\nonumber\\
+
&
\frac{3 (2 z-3) (3 z-2)}{2 z (1-z)}
\bigg] + O(\alpha_s^2),
\end{align}
where $z=(1-\cos\chi)/2$ and $C_F=4/3$ is the quadratic Casimir of the gauge group $SU(3)$. 
Already at the next-to-leading order, EEC is induced by final states with three and four particles. 
  
Despite numerous attempts (see Ref.~\cite{LEP2} and references therein), the analytical calculation of $O(\alpha_s^2)$ corrections 
to EEC is still an open problem. The main complications arise from the necessity to regularize infrared divergences of 
individual contributions to  \re{EEC}, and from the resulting complexity of the regularized multiparticle phase-space 
integrals. The Kinoshita-Lee-Nauenberg theorem warrants the absence of infrared singularities in the final result but it comes at the price of 
a  nontrivial cross-talk between contributions involving different numbers of particles in the final state.

A natural question is whether there exists an alternative approach to computing EEC that avoids the above problems. 
This is what we show in this work. First, we present such a framework and, second, we illustrate its power by calculating
EEC in $\mathcal{N}=4$  SYM, that can be viewed as a simplified version of QCD.   
As we show below, it shares many features with QCD as far as the properties of EEC are concerned.
 
Apart from gluons, $\mathcal{N}=4$ SYM also describes fermions and scalars, all in the adjoint 
representation of the $SU(N)$ gauge group. The theory has no scale and remains conformal for any value of the coupling 
$g_{\rm \scriptscriptstyle YM}$.
Since EEC is insensitive to the dynamics in the infrared, the latter property does not pose any problem and we can safely apply \re{EEC} to 
define EEC in $\mathcal{N}=4$ SYM to any order in perturbation theory. Furthermore, in the planar limit the theory is known to be 
dual to string theory on $AdS_5 \times S^5$ \cite{Mal}. This allows one to compute  EEC in planar $\mathcal{N}=4$ SYM at strong 
coupling.
 
 \medskip 
  
 \noindent \textbf{2. Energy flow correlations. }
The total cross section $\sigma_{\rm tot}$ is the simplest example of an observable for which the conventional approach based on scattering 
amplitudes proves inefficient. It is much more advantageous to 
use the optical theorem and compute $\sigma_{\rm tot}$ as the Fourier transform 
of two-point correlation functions of gauge invariant operators $O(x)$, 
\begin{align}\label{opt}
\sigma_{\rm tot}= \int d^4 x \, \e^{iqx}\vev{0| O^\dagger(x) O(0)|0}\,,
\end{align}
with $q^\mu = (Q, \vec{0})$ being the total momentum in the center-of-mass frame, $q^2>0$. Applying \re{opt}, we avoid infrared divergences at 
intermediate steps and the necessity to sum over all final states.

EEC \re{EEC} admits an equivalent representation analogous to \re{opt} in terms of the Wightman (non-time-ordered) correlation function
\begin{align}\label{EEC1}
\vev{\mathcal{E} (\vec{n}_1)   \mathcal{E} (\vec{n}_2)}_q= \int d^4 x \, {\rm e}^{i q \cdot x}
\vev{0| O^\dagger (x) \mathcal{E} (\vec{n}_1)   \mathcal{E} (\vec{n}_2) O (0)|0}
\end{align}
involving the so-called energy flow operators \cite{Flow,Hofman:2008ar,Belitsky:2013xxa}
\begin{align}\label{E}
\mathcal{E} (\vec{n})
=
  \int_{- \infty}^\infty d \tau  \lim_{r \to \infty} r^2 n^i T_{0i} (t = \tau+ r,r \vec{n})\,.
\end{align}
Here the stress-energy tensor $T_{\mu\nu}(t,\vec x)$ is placed infinitely far from the collision region and 
is integrated over the detector working time. The operator \re{E} describes a calorimeter and  has a simple physical interpretation: it measures 
the energy flux 
per unit solid angle in a given directions $\vec n$ (with $\vec n^2=1$). The product $\mathcal{E} (\vec{n}_1)   \mathcal{E} (\vec{n}_2)$
measures the correlation between energy flowing in the direction of $\vec n_1$ and $\vec n_2$. Then, EEC is given by
the correlation function \re{EEC1} averaged over the orientations $\vec n_1$ and $\vec n_2$, 
with the relative angle $\chi$ kept fixed
\begin{align}\label{EEC2}
{\rm EEC} = \int d \Omega_{1} d \Omega_{2}\,\delta(\vec n_1\cdot\vec n_2-\cos\chi) 
{\vev{\mathcal{E} (\vec{n}_1)   \mathcal{E} (\vec{n}_2)}_q \over Q^2\sigma_{\rm tot}}.
\end{align}
For a scalar source $O(x)$ and $q^\mu = (Q, \vec{0})$, the correlation function \re{EEC1} only depends on $(\vec n_1\cdot \vec n_2)$ so that the average \re{EEC2} becomes trivial.

Relations \re{opt}, \re{EEC1}  and \re{EEC2} rely on unitarity and the completeness of the asymptotic states,
$\sum_X \ket{X}\bra{X}=1$. They hold in a generic field theory, be it QCD or $\mathcal N=4$ SYM. To make use of \re{opt}  and 
\re{EEC1}, we have to specify the source $O(x)$ and find an efficient way of computing the Wightman correlation functions involving 
the energy flow operators. For $\e^+\e^--$annihilation in QCD, the operator $O(x)$ is 
given by the electroweak quark current. For the sake of simplicity, in $\mathcal N=4$ SYM  we choose 
it to be the simplest half-BPS operator of  dimension two, $O(x)=\tr[Z^2(x)]$, built from a complex scalar $Z(x)$.  

At weak coupling, the operator $\tr[Z^2(x)]$ produces out of the vacuum a pair
of complex scalars that decays into an arbitrary number of on-shell massless
$\mathcal N=4$ particles (gluons, gluinos and scalars).   EEC in 
$\mathcal N=4$ SYM receives contributions from the same type of Feynman diagrams as in QCD 
 with the only difference that the detected particles $a$ and $b$ can be
of different types. Another advantage of the choice of half-BPS operators is that the two-point correlation
function $\vev{0|O^\dagger(x) O(0)|0}$ is protected in $\mathcal N=4$ SYM from quantum corrections and is given by its Born
approximation. Together with \re{opt} this leads to $\sigma_{\rm tot} = (N^2-1)/(4\pi)$ which is 
valid in $\mathcal N=4$ SYM for arbitrary coupling. Unlike $\sigma_{\rm tot}$,  EEC is fixed up to an arbitrary
function $F(z;a)$ of the angle $\chi$ and the 't Hooft coupling $a=g^2_{\scriptscriptstyle\rm YM} N/(4\pi^2)$
\begin{align}\label{F-def}
{\rm EEC}_{\rm \scriptscriptstyle \mathcal N=4} =  \frac{ F(z;a)}{4z^2(1-z)}\,,\qquad   z= \sin^2 (\chi/2)
.
\end{align}
Here $0< z<1$ and the prefactor is chosen for convenience.
EEC is expected to be a regular positive-definite function of $z$, normalized as
$\int_0^1 d z \, {\rm EEC}(z)=1/2$. 
 
In close analogy with the QCD result \re{1loopQCD}, the weak-coupling expansion of $F(z,a)$ starts at order $O(a)$. The lowest order term 
comes from the transition of the operator $\tr[Z^2(x)]$ into three-particle states (two scalars plus a gluon and a scalar plus a pair of 
gluinos). It reads \cite{Ede}
\begin{align}
\label{N4EEC1loop}
{\rm EEC}_{\rm \scriptscriptstyle \mathcal N=4} =   \frac{a}{4 z^3}  
{{z\over 1-z} } \ln {1\over 1-z} + O(a^2)
\, .
\end{align}
Comparing this relation with  \re{1loopQCD}, we observe that (up to the redefinition of the coupling $a\to \alpha_s C_F/\pi$) 
EEC in QCD and in $\mathcal{N}=4$ SYM have identical asymptotics for $z\to 1$. Also, both EECs exhibit the 
same $O(1/z)$ behavior for $z\to 0$, but the coefficient driving this asymptotics is different. For $z\to 1$, or equivalently $\chi\to \pi$, 
EEC measures back-to-back correlations and its asymptotic behavior is governed by the emission of soft and collinear particles.
Their contribution can be analyzed using the semiclassical approximation and is known to be universal, independent of the choice of the 
source $O(x)$ in both theories.  On the other hand, the $z\to 0$ asymptotics describes the correlation between particles with almost
aligned momenta. It is driven by the collinear branching of particles in the final state and is sensitive to the particle content
of the theory. 

\medskip
 
\noindent \textbf{3. Method and result.}
Trying to compute the next-to-leading $O(a^2)$ correction to \re{N4EEC1loop} using the conventional technique based on scattering
amplitudes, we encounter the same complications as in QCD. They can be overcome by employing 
\re{EEC1} and applying the formalism developed in Refs.~\cite{Belitsky:2013xxa}. It allows us to avoid  
infrared divergences and provides a framework that preserves all symmetries of the theory at every step of the calculation. 

Replacing $\mathcal{E}$ in \re{EEC1} by its definition \re{E}, we see that EEC is related to a multi-fold integral of the 
four-point Wightman correlation function $\vev{O^\dagger T_{\mu \nu} T_{\rho \sigma} O}$ involving two half-BPS operators and two 
stress tensors. In $\mathcal{N}=4$ SYM it has the following remarkable properties. 
First, the superconformal symmetry relates it to the one built from four scalar half-BPS operators, thus
eliminating the complication of dealing with Lorentz indices. Second, in Euclidean space, the latter correlation 
function is defined by a single function $\Phi (u, v; a)$ of the conformal cross ratios $u$ and $v$ and admits  the Mellin representation 
\begin{align}\label{M}
\Phi(u,v; a) =  \int_{-\delta-i\infty}^{-\delta+i\infty} {dj_1 dj_2 \over (2\pi i)^2} M(j_1,j_2; a) u^{j_1} v^{j_2}\, .
\end{align}
The Mellin amplitude $M(j_1,j_2; a)$ is known both at weak and strong coupling in planar  $\mathcal{N}=4$ SYM. 
The integral in \re{M} goes along the imaginary axis with $0<\delta<1/2$.

Following  the L\"uscher-Mack procedure, we can analytically continue \re{M} to obtain the four-point Wightman 
functions in Minkowski space \cite{Mack}. Finally, performing the  
limit and integration as 
indicated in \re{E} we obtain from \re{EEC1} and \re{F-def} the following representation for $F(z;a)$ \cite{Belitsky:2013xxa}
\begin{align}\label{convolution}
 F(z; a) &=  \int {dj_1 dj_2 \over (2\pi i)^2} M(j_1,j_2; a) K_{\rm \scriptscriptstyle EEC} (j_1,j_2;z)
\, .
\end{align}
The dependence on the coupling constant resides in the Mellin amplitude, while the $z$-dependence enters only
the kernel encoding the information about the calorimeters,
\begin{align}\label{K}
K_{\rm \scriptscriptstyle EEC} 
(j_1,j_2;z) &= {2\, \Gamma(1-j_1-j_2)  \lr{\displaystyle z\over 1- \displaystyle z}^{-j_1-j_2} \over \Gamma(j_1+j_2)[\Gamma(1-j_1)\Gamma(1-j_2)]^2}\, .
\end{align}
Thus, Eq.~\re{convolution} yields EEC in $\mathcal{N}=4$ SYM as the convolution of two well-defined functions. 

To compute \re{convolution} at next-to-leading order, we have to expand the Mellin amplitude to order $O(a^2)$. To this end, we 
make use of the known two-loop result  \cite{Kor,Eden:2012tu} for the function \re{M}  
\begin{align*} 
&
\Phi(u,v; a) =  a \,\Phi^{(1)}(u,v) + a^2 \bigg\{ \frac12(1+u+v) \left[\Phi^{(1)}(u,v) \right]^2 
\\
&+ 2 \left[  \Phi^{(2)}(u,v) 
+ \frac1{u} \Phi^{(2)}(v/u,1/u) +  \frac1{v} \Phi^{(2)}(1/v,u/v) \right]
\!\!\bigg\},
\end{align*}
in terms of  Euclidean scalar box integrals 
$\Phi^{(1)}$ and  $\Phi^{(2)}$
whose explicit form can be found in Refs.\ \cite{Kor,Eden:2012tu,DavUsy}.
All that we need for our purposes is their Mellin transforms \cite{DavUsy}
\begin{align*} 
&M^{(1)} (j_1,j_2) = -{1\over 4} \left[ \Gamma(-j_1) \Gamma(-j_2)\Gamma(1+j_1+j_2)\right]^2
\, ,
\end{align*}
\begin{align*} 
&M^{(2)} (j_1,j_2) =-{1\over 4}  \Gamma(-j_1) \Gamma(-j_2)\Gamma(1+j_1+j_2)\int {dj_1' dj_2' \over (2\pi i)^2}
 \\
&
\times  {\Gamma(j_1'-j_1)\Gamma(j_2'-j_2) \Gamma(1+j_1+j_2-j_1'-j_2') \over \Gamma(1-j_1')\Gamma(1-j_2') \Gamma(1+j_1'+j_2')}
M^{(1)} (j_1',j_2').
\end{align*}
Using these relations it is straightforward to work out the expression for the Mellin amplitude in \re{M},
\begin{align}\notag\label{M2}
M =&  a M^{(1)} (j_1,j_2) + a^2 \left[\ft12 \widetilde M^{(2)}(j_1,j_2)  +\widetilde M^{(2)}(j_1,j_2-1) \right.
\\
& \left. + 2M^{(2)} (j_1,j_2) + 4M^{(2)} (j_1,-1-j_1-j_2)\right],
\end{align}
where $\widetilde M^{(2)}(j_1,j_2)$ is the Mellin amplitude for $[\Phi^{(1)}]^2$,
\begin{align*}
\widetilde M^{(2)} =  \int {dj_1' dj_2' \over (2\pi i)^2} M^{(1)} (j_1-j_1',j_2-j_2')M^{(1)} (j_1',j_2').
\end{align*}
Substituting \re{K} and \re{M2} into \re{convolution} we find $F(z;a)$ as a sum of nested Mellin integrals which can be computed using standard techniques. 

To save space we present the final result of our calculation
and defer the details to a separate publication. It is convenient to write EEC as
\begin{align}\label{F-res}
F (z;a) = a F_1(z) + a^2 \left[ (1-z)F_2(z) + F_3(z)\right] ,
\end{align}
where $0<z<1$ and $F_w(z)$ stands for a linear combination of functions of homogenous weight $w=1,2,3$ specified below. To lowest order
in the coupling we have
\begin{align}
F_1(z) = -  \ln(1-z)
\end{align}
in agreement with \re{N4EEC1loop}. At next-to-leading order, the functions $F_2(z)$ and $F_3(z)$ take the form
\begin{widetext}
\begin{align} \notag 
&F_{2}(z)
= 4\sqrt{z} \left[{\rm Li}_2 \left(-\sqrt{z} \right)   
-  {\rm Li}_2 \left(\sqrt{z}\right) +{ \ln z  \over 2}\ln \left({1 + \sqrt{z} \over 1 - \sqrt{z} }\right)  \right] 
+ (1+z) \left[2 {\rm Li}_2 (z) +\ln^2(1-z)\right] + 2 \ln (1-z) \ln \left({z \over 1-z}\right) 
+ z {\pi^2 \over 3}, \\
 \notag \label{weightthree}
&F_{3}(z)
=\frac14\bigg\{(1-z)(1+2 z) \left[\ln^2 \left({1+ \sqrt{z} \over 1 - \sqrt{z}} \right) \ln \left({1 - z \over z}\right) 
-8{\rm Li}_3 \left({\sqrt{z} \over \sqrt{z} - 1} \right)   - 8 {\rm Li}_3 \left({\sqrt{z} \over \sqrt{z} + 1} \right)   \right] -4 (z-4) {\rm Li}_3 (z) \\ \nonumber
&\qquad\quad
+ 6 (3 + 3 z - 4 z^2) {\rm Li}_3 \left({z \over z - 1} \right) - 2 z (1+4 z) \zeta_3+2 \left[2(2 z^2-z-2)\ln(1-z) +(3-4z) z \ln z \right] {\rm Li}_2 (z) \\ 
&\qquad\quad
+{1 \over 3} \ln^2 \left(1-z \right) \left[4(3 z^2 -2 z -1) \ln (1-z) +3 (3-4 z) z \ln z \right] + {\pi^2 \over 3} \left[ 2 z^2 \ln z - (2 z^2 + z - 2) \ln(1-z) \right]\bigg\}.
\end{align}
\end{widetext}
Both functions are given by sums of  basis  functions of  weight two $\{{\rm Li}_2, \ln \ln ,\pi^2\}$ and 
weight three $\{{\rm Li}_3, {\rm Li}_2 \ln, \ln \ln \ln,\pi^2\ln,\zeta_3\}$, respectively, with nontrivial arguments. The unusual features of 
$F_2(z)$ and $F_3 (z)$ is that the basis functions come with factors of $\sqrt{z}$ and $z$ for the former and prefactors being at most 
quadratic in $z$ for the latter. Eq.\ \re{weightthree} represents the main result of this work. 

The following comments are in order. Even though the functions $F_2(z)$ and $F_3(z)$ depend on  $\sqrt{z}$, they are manifestly invariant 
under $\sqrt{z}\to -\sqrt{z}$, so that their expansion near $z=0$ runs in integer powers of $z$  as described below. The  cross sections 
$d\sigma_{V\to {a+b}+X}$ entering \re{EEC} are given by (absolute value squared of) scattering amplitudes which are known to have 
homogenous weights in planar $\mathcal N=4$ SYM at weak 
coupling. As we see, this property is lost for EEC after the phase space integration in the right-hand side of  \re{EEC}. Nevertheless, it is 
restored for $z\to 1$ as we show below.
 
 \medskip
 
\noindent \textbf{4.~End-point asymptotics.}
Let us examine  the asym\-pto\-tics of EEC close to the end points $z=0$ and $z=1$.

As we mentioned above, in the back-to-back kinematics $z\to 1$ (or $\chi \to \pi$), EEC receives large perturbative (Sudakov) corrections from the 
emission of soft and collinear particles. We find from \re{F-res} (with $y\equiv 1-z\to 0$)
\begin{align}
\label{LargezF2}
F(z;a) \stackrel{z \to 1}{\sim} -{a} \ln y +{a^2\over 2}   \left[  \ln^3 y +{\pi^2\over 2}\ln y +\zeta_3   \right] ,
\end{align}
up to terms suppressed by powers of $y$. The logarithmically enhanced corrections
$a^k \ln^n y$ can be resummed to all orders in the coupling in very much the same way as in QCD \cite{Collins:1981uk} yielding 
\begin{align}
\label{SudakovFN=4}
 F (z; a)
& \stackrel{z \to 1}{\sim}  \frac12H (a)
\int_0^\infty db \, b \, J_0 (b) S (b^2/y; a)\,.
\end{align}
Here $J_0(b)$ is a Bessel function and $S (b^2/y; a)$ is the Sudakov form factor
\begin{align*}
S = \exp\left[
- \frac{1}{2} \Gamma_{\rm cusp} (a) \ln^2 \lr{ \frac{b^2}{y b_0^2} } - \Gamma (a) \ln \lr{ \frac{b^2}{y b_0^2} }
\right]
\end{align*}
(with $b_0=2\e^{-\gamma_{\rm E}}$). Its dependence on the coupling constant is encoded in two functions, the cusp ($\Gamma_{\rm cusp}$) 
and collinear ($\Gamma$) anomalous dimensions. At weak coupling, to two loop order, we have $\Gamma_{\rm cusp} (a) = a - \ft12 \zeta_2 a^2$ 
and $\Gamma (a) = - \ft32 \zeta_3 a^2$. Both functions are known from integrability in planar $\mathcal{N} = 4$ SYM for any coupling.
Finally, $H (a)$ is the so-called coefficient function that accounts for the emission of hard gluons. It only depends on the coupling constant and 
is needed to incorporate subleading logarithms.  Expanding \re{SudakovFN=4} to order $O(a^2)$ we reproduce \re{LargezF2} and obtain 
$H (a) = 1 - \zeta_2 a$. Note that the perturbative correction to $H(a)$ has homogeneous transcedentality. The same applies to all functions of 
the coupling entering \re{SudakovFN=4} and hence to $F(z;a)$ itself. Indeed, examining \re{F-res} we observe that the contribution of the weight-two 
function $F_2(z)$ is suppressed for $z\to 1$ whereas $F_3(z)$ reduces to a sum of weight-three functions with rational coefficients. Moreover, we 
anticipate that for $z\to 1$ the function $F(z;a)$ computed in $\mathcal N=4$ SYM describes the maximally transcendental part of the analogous QCD 
expression. We verified it to two-loop accuracy making use of the available results \cite{Dokshitzer:1999sh}.

Let us now turn to the analysis of the opposite asymptotic limit $z \to 0$  (or $\chi\to 0$). It corresponds to the physical situation where the 
calorimeters measure nearly collinear particles.
We find from \re{F-res} that in this limit EEC gets enhanced by a $\ln z$,
\begin{align}
\label{SmallzF2}
F(z;a) \stackrel{z \to 0}{\sim} {a z} \left[1+ a \left( \ln z - \ft12 \zeta_3 + \zeta_2 - 3 \right) \right].
\end{align}
Notice that, in spite of the presence of $\sqrt{z}$ in the two-loop result  \re{weightthree},
the function in \re{SmallzF2} has an expansion in integer positive powers of $z$. All log-enhanced corrections to $F(z;a)$ of the form $a(a\ln z)^k$ 
can be resummed using the ``jet calculus'' \cite{Konishi:1979cb}.
In this approach, the leading $z\to 0$ asymptotics of EEC is
determined by a partonic cascade in which the scalar particle with virtuality $Q^2$ created from the vacuum by the source (half-BPS operator) 
fragments into a pair of partons $a$ and $b$ with a small invariant mass $S_{ab} = 4 E_{a} E_{b} \sin^2(\chi/2) \sim Q^2 z$. EEC is then
given by the total probability of this transition weighted with the energy of the detected particles,
\begin{align}\label{z0}
F(z;a) \stackrel{z \to 0}{\sim} {a z}  (Q^2/S_{ab})^{-\gamma_T(3)} =  {a} z^{1+a + O(a^2)},
\end{align}
where $\gamma_T(S) = a \sum_{k=1}^{S-2} 1/k + O(a^2)$ is the twist-two time-like anomalous dimension of spin $S$. 

As follows from \re{SudakovFN=4} and \re{z0},  the resummation weakens the singularity of $F(z;a)$ at $z=1$ and $z=0$, respectively, so 
that the energy-energy correlation \re{F-def} becomes integrable at the end points. The resummation formulas \re{SudakovFN=4} and \re{z0}
can be combined with \re{weightthree} to provide a definite prediction for EEC for $0\le z\le 1$ in the next-to-leading logarithmic approximation. 
We find that the obtained EEC has a shape which is remarkably similar to the one in QCD \cite{LEP2}. Going from one to two loops, we observe 
that EEC flattens. This is in qualitative agreement with the prediction for EEC in planar $\mathcal N=4$ SYM at strong coupling 
\cite{Hofman:2008ar,Belitsky:2013xxa}
\begin{align*}
{\rm EEC}_{\rm \scriptscriptstyle \mathcal N=4} \stackrel{a\to\infty}{\sim}  \frac12 \left[1+ a^{-1}\left(1-6z(1-z)\right)+O(a^{-3/2}) \right].
\end{align*}
It will be challenging to find a function that interpolates EEC between weak and strong coupling. Finally, we believe that our present analysis
can help in advancing analytic calculations of EEC in QCD. It is worth mentioning that \re{weightthree} involves some of the transcendental 
functions that also appear in the two-loop result for the quarks (or $n_f$ dependent) contribution to EEC in QCD \cite{DS}.
 
We would like to thank L.~Dixon, J.~Henn and D.~Kosower for helpful discussions. We are especially grateful to J.~Henn for his generous help in 
carrying out Mellin integration. The work of A.B. and A.Z. was supported by the U.S. National Science Foundation under the
grants PHY-1068286 and PHY-0756966, respectively. G.K. and E.S. acknowledge partial support by the French
National Agency for Research  under contract BLANC-SIMI-4-2011.

\end{document}